\documentclass[a4paper,fleqn]{cas-sc}
\usepackage{relsize}
\usepackage[authoryear]{natbib}
\usepackage{hyperref}
\def\tsc#1{\csdef{#1}{\textsc{\lowercase{#1}}\xspace}}
\tsc{WGM}
\tsc{QE}
\tsc{EP}
\tsc{PMS}
\tsc{BEC}
\tsc{DE}

\begin{document}
\let\WriteBookmarks\relax
\def\floatpagepagefraction{1}
\def\textpagefraction{.001}
\shorttitle{Estimating the $^{56}$Ni mass}
\shortauthors{J. Gaba et~al.}

\title [mode = title]{How accurate are current $^{56}$Ni mass estimates in Type Ia Supernovae?}

\author[1]{Jagriti Gaba}[
                        orcid=0009-0003-2207-2689]
\ead{jagritidang@gmail.com}

\affiliation[1]{organization={School of Engineering \& Sciences, GD Goenka University},
                addressline={Sohna}, 
                city={Gurugram},
                postcode={122103}, 
                state={Haryana},
                country={India}}

\author[2]{Rahul Kumar Thakur}[
]
\ead{thakurr58@gmail.com}
\affiliation[2]{organization={Avantika University},
                city={Ujjain},
                postcode={456006}, 
                state={Madhya Pradesh},
                country={India}}

\author[1]{Naresh Sharma} [%
   orcid=0000-0001-8398-927X]
\ead{naresh.sharma2006@gmail.com}

\author[1]{Dinkar Verma} [%
   orcid=0000-0002-8744-5891]
\cortext[cor1]{Corresponding author}
\cormark[1]
\ead{verdinkar@gmail.com} 

\author[1]{Shashikant Gupta}[%
orcid=0000-0003-2923-9245]
\cormark[2]
\ead{shashikantgupta.astro@gmail.com}
\cortext[cor2]{Principal corresponding author}
\maketitle
\begin{abstract}
The diversity of type Ia supernovae (SNe Ia) has become increasingly apparent with the rapid growth in observational data. Understanding the explosion mechanism of SNe Ia is crucial for their cosmological calibration and for advancing our knowledge of stellar physics. The estimation of $^{56}$Ni mass produced in these events is key to elucidating their explosion mechanism. This study compares two methods of $^{56}$Ni mass estimation.
We first examine the relationship between peak luminosity and the second maximum in near-infrared (NIR) bands using observations of 18 nearby SNe Ia. Based on this relationship, we estimate the Ni mass for a set of nine well-observed SNe Ia using the Arnett rule. Additionally, we estimate the $^{56}$Ni mass using bolometric light curves of these SNe through energy conservation arguments.
A comparison of these two estimation methods using Student's t-test reveals no statistically significant differences between the estimates. This finding suggests that both methods provide robust estimates of Ni mass in SNe Ia.

\end{abstract}

\begin{keywords}
Stars \sep Supernova \sep White Dwarf
\end{keywords}

\section{Introduction}
Type Ia Supernovae (SNe Ia) are stellar explosions that occur in binary star systems where one of the stars is a white dwarf (WD). These cosmic events, triggered by the thermonuclear detonation of the white dwarf, have become cornerstone objects in modern cosmology due to their role as standardizable candles. The importance of SNe Ia in cosmology stems from their remarkable uniformity in light curve shapes and spectroscopic properties, as well as the correlation between their peak luminosity and decline rate. With a peak absolute magnitude of approximately $M \approx -19$, these celestial beacons are bright enough to be detected at cosmological scales. 
The standard model of cosmology, along with other astronomical probes \citep{spergel2003first}, relies on the observations of SNe Ia \citep{riess1998observational, perlmutter1998discovery}. The standard model predicts a cosmic composition of approximately $5\%$ ordinary baryonic matter, $25\%$ pressure-less dark matter, and $70\%$ dark energy – a mysterious component with negative pressure. Thus, SNe Ia helps us understand a variety of astrophysical phenomena, ranging from cosmology to stellar evolution, as well as the secrets of the universe's past, present, and future. 
The physics of Type Ia supernovae (SNe Ia) remains poorly understood, presenting challenges both theoretically and observationally. The brightness and other light curve characteristics of these supernovae can vary depending on the properties of their host galaxies. Additionally, there may be intrinsic diversity and systematic issues within the observed data
\citep{gupta2010direction, gupta2014high, moreno2016dependence, pruzhinskaya2020dependence, ponder2021type, arima2021intrinsic, pierel2021understanding, thakur2023investigating}.
Issues related to accretion rates, the effect of rotation, and the composition of WD also require further investigation \citep{gilfanov2010upper, saruwatari2010effects, pfannes2010thermonuclear, yungelson2010evolution, wang2014evolution, ghosh2017differentially, neunteufel2017helium, wang2017he, wang2018mass, fink2018thermonuclear, schwab2019residual, kumar2023accreting}. Recent reviews on Type Ia supernovae in binary stellar systems can be found in works by
\cite{wang2018mass, liu2023type} 
Thus, to ensure the validity of the standard cosmological model, it is crucial to comprehend the explosion mechanism and accurately determine the physical parameters of SNe Ia. 
Two primary models attempt to explain the explosion mechanism of SNe Ia:
\begin{itemize}
\item The single-degenerate model: A white dwarf accretes matter from a companion star until it reaches a critical mass, triggering a thermonuclear explosion.
\item The double-degenerate model: Two white dwarfs merge, leading to a detonation or deflagration event due to gravitational interactions. 
\end{itemize}
While these models provide valuable insights, SNe Ia's precise explosion mechanism is still uncertain.
One of the key uncertainties lies in the mass of $^{56}$Ni ($M_{^{56}\mathrm{Ni}}$) produced during SNe Ia explosions, with estimates ranging from 0.4 to 1.2 $M_{\odot}$ \citep{ohlmann2014white, lach2022type}. This uncertainty has significant implications for our understanding of the explosion mechanism and the resulting brightness of the supernova. Additionally, the production of intermediate-mass elements (IMEs) such as silicon, magnesium, and sulfur in SNe Ia is not yet fully understood, impacting our knowledge of nucleosynthesis and galactic chemical enrichment \citep{seitenzahl2013solar, reeves2023dependence}. 
Accurate determination of the $^{56}$Ni mass is crucial as it directly influences the luminosity of SNe Ia, which is the foundation for their use as standard candles in cosmological distance measurements. Uncertainties in progenitor systems and explosion mechanisms contribute to the challenge of precisely estimating $M_{^{56}\mathrm{Ni}}$ \citep{ dado2015analytical, liu2023type}. Addressing these issues is essential for understanding stellar evolution and cosmological theories.

In this work, we focus on estimating the mass of $^{56}$Ni produced in Type Ia supernovae using various methods, including analyzing light curves and spectra. To estimate the mass of $^{56}$Ni produced in Type Ia supernovae, researchers typically analyze the light curves and spectra of the explosion. One popular approach involves using theoretical models that incorporate the radioactive decay chain of $^{56}$Ni to $^{56}$Co, and finally to $^{56}$Fe. By fitting these models to observed light curves, astronomers can infer the initial mass of $^{56}$Ni present during the explosion. This mass estimation is crucial because the intense combustion of carbon and oxygen ultimately produces $^{56}$Ni, which then decays to $^{56}$Co and $^{56}$Fe through beta decay. The rate of these radioactive decays provides valuable information about the initial amount of $^{56}$Ni synthesized, allowing us to determine the overall mass of nickel produced in the supernova.

Variations in the bolometric luminosity of SNe Ia introduce uncertainties in the derived physical parameters of the explosions, particularly the synthesized $M_{^{56}\mathrm{Ni}}$ and total ejecta mass. These variations have been observed in previous studies \citep{stritzinger2006consistent, scalzo2014type}. Additionally, data from these studies suggest that the peak luminosity exhibits non-uniformity in the optical range but displays a consistent brightness distribution in the near-infrared (NIR) wavelengths ($900\, nm <\lambda < 2000\, nm$). Interestingly, recent research \citep{dhawan2015near, dhawan2016reddening} has revealed strong correlations between the timing of the second maximum in the NIR band and the optical light-curve shape. 
Furthermore, the bolometric luminosity peak is found to be related to the timing of the second maximum ($t_{2}$) corresponding to the NIR band. This relation, along with  Arnett's rule \citep{arnett1982type} can be used to derive the $^{56}$Ni masses.
In another study by  \citep{wygoda2019type}, the energy conservation method is employed to infer the mass of $^{56}$Ni based on accurately constructed bolometric light curves. This method explores the physical relationship between gamma-ray escape time $t_{0}$ and $^{56}$Ni, termed the ``bolometric WLR," across different brightness levels.

In this paper, we extend the application of Arnett's rule \citep{arnett1982type} to a new sample of SNe Ia and investigate the correlation between $L_{\mathrm{max}}$ and $t_{2}$. We utilize a sample of 18 and 9 nearby SNe Ia (described in Section 2) and employ two different methods to derive $M_{^{56}\mathrm{Ni}}$, as detailed in Section 3. The derived masses of $^{56}$Ni for our sample are presented in Section 4. Furthermore, we explore the application of the energy conservation method utilizing bolometric light curves from existing SNe Ia literature. A comparative analysis between the derived $M_{^{56}\mathrm{Ni}}$ values obtained through the energy conservation method and those derived using Arnett's rule \citep{arnett1982type} methods offers valuable insights into the accuracy and potential limitations of the latter. The findings have been summarized in Section 5.

\section{Data}

Two separate data samples of SNe Ia have been used in our analysis. The first sample comprises 18 well-sampled nearby SNe Ia with NIR data in different bands. These SNe have been taken from Table 1 of \cite{dhawan2016reddening}. The main criterion for selecting the sample was the availability of the secondary peak in the NIR data along with the bolometric luminosity $L_{\mathrm{max}}$. It has been used to determine a relation between the secondary maxima and the peak luminosity.

The second data set of our analysis consists of nine SNe Ia that have been specifically chosen for their well-defined NIR light curves with secondary maxima listed in Table 1. We have used these SNe Ia to calculate the nickel mass. The additional observations, such as light-curve shape parameters, and rise time, are crucial for further analysis in our study. Each SNe contributes unique characteristics that will aid in understanding Type Ia supernovae behavior and properties.

\begin{table}
\caption{Sample of 9 SNe Ia along with the light curve shape parameter ($\Delta m_{15}$) and the time of second maximum ($t_2$).} 
\label{table:data}
\begin{tabular*}{\tblwidth}{@{}LCCCC@{}}
\toprule
SNe & $\Delta m_{15}$ &  $t_2$ & Reference for $\Delta m_{15}$ and $t_2$ & Reference for Bolometric Light Curve\\ 
\midrule
SN2005el & 1.4 & 24.6 & \cite{dhawan2015near} & \cite{scalzo2014type}\\ 
SN1998bu & 1.01 & 29.84 & \cite{dhawan2015near} & \cite{contardo2000epochs}\\ 
SN2007on & 1.65 & 18.2 & \cite{dhawan2015near} & \cite{phillips2012near}\\
SN2006X & 1.09 & 28.19 & \cite{dhawan2015near} &  \cite{wang2008optical}\\
SN2005cf & 1.12 & 30.53 & \cite{pastorello2007esc} &   \cite{pastorello2007esc}\\
SN2003du &  1.02 & 32.59 & \cite{stanishev2007sn} & \cite{stanishev2007sn}\\
SN2011fe & 1.2 & 30 & \cite{dhawan2015near} &  \cite{dado2015analytical}\\
SN2002bo & 1.12 & 28.9 & \cite{dhawan2015near} &  \cite{benetti2004supernova}\\
SN2000E & 0.99 & 31.85 & \cite{dhawan2015near} &  \cite{biscardi2012slowly}\\
\bottomrule
\end{tabular*}
\end{table}

\section{Methodology}
\label{sec:methodology}

Understanding the decay chain of $^{56}$Ni to $^{56}$Co to $^{56}$Fe ($^{56}\rm{Ni} \rightarrow \,  ^{56}\rm{Co}  \rightarrow \, ^{56}\rm{Fe}$) is crucial for estimating the mass of $^{56}$Ni produced in Type Ia Supernovae (SNe Ia). The decay chain involves the radioactive transformation of $^{56}$Ni into $^{56}$Co through electron capture decay, followed by the decay of $^{56}$Co into a stable iron isotope, $^{56}$Fe, via positron decay or electron capture \citep{nadyozhin1994compilation}. These decay processes are described by the following equations:

\begin{equation}
^{56}\rm{Ni} \rightarrow \rm{^{56}{Co}} + \gamma + \nu_e\,,
\label{eq:decaychain1}
\end{equation}

\begin{equation}
    ^{56}\rm{Co} \rightarrow \rm{^{56}{Fe}} + e^+ + \gamma + \nu_e \,\, (Positron \, Decay),
    \label{eq:decaychain2}
\end{equation}
\begin{equation}
    ^{56}\rm{Co} \rightarrow \rm{^{56}{Fe}} + \gamma + \nu_e \,\, (Electron \, Capture).
    \label{eq:decaychain3}
\end{equation}

 In the following subsections, we present a theoretical framework to employ the two distinct methods to estimate the mass of $^{56}$Ni for each SNe Ia and compare the results. The first method is based on Arnett's rule \citep{arnett1982type}, a well-established technique that relates the peak luminosity of SNe Ia to the energy deposition rate from radioactive decay within the expanding ejecta. The second method utilizes energy conservation and bolometric UVOIR light curves to bypass the complexities of radiation transfer mechanisms \citep{wygoda2019type}.


\subsection{Estimating $M_{^{56}\mathrm{Ni}}$ Using Arnett's Rule}
\label{sec:arnett}
\subsubsection{The Arnett's Rule}
According to Arnett's rule, the energy deposition rate from radioactive decay inside the expanding ejecta determines the SNe's peak luminosity \citep{arnett1982type}. For Type Ia supernovae, the peak bolometric luminosity, $L_{\mathrm{max}}$, can be described by the following expression:
\begin{equation}
L_{max}(t_R) = \alpha E_{^{56}\rm{Ni}}(t_R),
\end{equation}
where $E_{^{56}\rm{Ni}}(t_R)$ represents the rate of energy input from the decays of $^{56}$Ni and $^{56}$Co at the time of maximum luminosity, $t_R$ is the rise time to this bolometric maximum, and $\alpha$ is a parameter that accounts for deviations from Arnett’s Rule \citep{stritzinger2006consistent}.

The energy output in erg/s due to the radioactive decay of 1 $M_\odot$ of $^{56}$Ni is given by:
\begin{equation}
\mathlarger{\mathlarger{\epsilon}}_{\mathrm{Ni}}(t_R, 1M_\odot) = (6.45 \times 10^{43} e^{-t_R/8.8} + 1.45 \times 10^{43} e^{-t_R/111.3}) \, .
\label{eq:gammaenergy}
\end{equation}
In this equation, the terms $6.45 \times 10^{43} e^{-t_R/8.8}$ and $1.45 \times 10^{43} e^{-t_R/111.3}$ correspond to the energy contributions from the decays of $^{56}$Ni and $^{56}$Co, respectively, at the rise time $t_R$. The exponential factors represent the decay lifetimes of $^{56}$Ni ($8.8$ days) and $^{56}$Co ($111.3$ days).
The B band rise time, $t_{R,B},$ can be estimated from $\Delta m_{15}$ a directly observed quantity \citep{scalzo2014type}:
\begin{equation}
t_{R,B} = 17.5 - 5(\Delta m_{15} - 1.1) \, .
\label{eq:tr}
\end{equation}
Given this relation, the mass of $^{56}$Ni can be estimated using the following expression:
\begin{equation}
\frac {M_{^{56}\mathrm{Ni}}}{M_\odot} = \frac{L_{max}}{\mathlarger{\mathlarger{\epsilon}}_{\mathrm{Ni}}(t_R, 1M_\odot)} \, .
\end{equation}
It has been found that the timing of the second maximum in NIR bands, \( t_2 \), is strongly correlated with \( L_{\text{max}} \) \citep{dhawan2016reddening}. This correlation can be used as a proxy whenever direct measurements of the peak bolometric luminosity, \( L_{\text{max}} \), are not available. 

\subsubsection{Secondary Maxima in NIR Light Curves and its correlation with $L_{\mathrm{max}}$}
The NIR light curves of SNe Ia exhibit a secondary maximum, \( t_2 \), which arises due to changes in the ionization state of iron (Fe) in the supernova \citep{kasen2006secondary, dhawan2017near}. It is particularly interesting that the NIR light curve parameters are correlated with well-known optical light curve shape parameters, such as \(\Delta m_{15}\). Specifically, the phase of the second maximum, \( t_2 \), in the J and Y bands shows a significant correlation with \(\Delta m_{15}\):
\begin{equation}
t_2(J) = (-20.3 \pm 0.9)  \Delta m_{15}  + ( 5.17 \pm 1.1) \, ,
\label{eq:t2jdm}
\end{equation}
and
\begin{equation}
t_2(Y) = (-20.6 \pm 1.0) \Delta m_{15}  + ( 53.6 \pm 1.2) \, .
\label{eq:t2ydm}
\end{equation}
In the absence of direct observations of \( t_2 \) in the NIR bands, one can use Eqs.~\ref{eq:t2jdm} and \ref{eq:t2ydm} to estimate \( t_2 \) from \(\Delta m_{15}\).

A simple linear relation between \( L_{\text{max}} \) and \( t_2 \) is often employed \citep{dhawan2016reddening}:
\begin{equation}
L_{\mathrm{max}} = a  t_{2} + b \, ,
\label{eq:lin}
\end{equation}
where \( a \) and \( b \) are fit parameters. This relationship has been demonstrated to exist in different NIR bands. To verify the adequacy of the linear fit, we also test a quadratic fit to the data using the relation:
\begin{equation}
L_{\mathrm{max}} = c  t_{2}^2 + a  t_{2} +b
\label{eq:cubic}
\end{equation}
We utilize data from 18 SNe Ia as presented in \cite{dhawan2016reddening} to evaluate these fits.

\subsubsection{Fixed vs Individual Rise Time}
The rise time for SNe Ia, \( t_R \), can be derived from \(\Delta m_{15}\) using Eq.~\ref{eq:tr}. Nevertheless, it has been observed that the rise times for most SNe Ia cluster around a narrow range near 19 days. Therefore, we also consider a fixed rise time of \( 19 \pm 3 \) days for the calculation of $M_{^{56}\mathrm{Ni}}$. The peak luminosity then becomes:
\begin{equation}
L_{max}= (2.0 \pm  0.3) \times 10^{43} {\frac{M_{^{56}\mathrm{Ni}}}{M_\odot}} \,  \mathrm{ erg/s} \, .
\end{equation}

\subsection{Energy Conservation Method}
\label{sec:ECM}
This method relies on the conservation of energy in the explosion and has been borrowed from\cite{wygoda2019type}. In SNe Ia, gamma rays and positrons are emitted during the radioactive decay of $^{56}$Ni and $^{56}$Co (Eq.~\ref{eq:decaychain1}--\ref{eq:decaychain3}). These particles travel through the supernova ejecta, depositing part of their energy into the surrounding plasma at a rate denoted as $R_{\mathrm{dep}}$. Initially, the ejecta is dense and opaque, ensuring that all gamma-ray energy is trapped and deposited within the ejecta. However, as the ejecta expands and becomes optically thin, a fraction of the gamma rays ($f_{\gamma}$) escapes without depositing their energy. The fraction of gamma rays depositing their energy in the ejecta is a function of time and can be described by:
\begin{equation}
    f_{\gamma}(t) = 1 - \exp{(-t_0^2/t^2)} \, ,
\end{equation}
where $t_0$ is the gamma-ray escape time, a characteristic time scale that remains constant over time $t$. The energy deposition rate, $R_{\mathrm{dep}}(t)$, accounts for both the gamma-ray and positron contributions. It can be expressed as:
\begin{equation}
R_{\mathrm{dep}}(t) = f_{\gamma}(t) R_{\gamma} + R_{\mathrm{dep},\, \mathrm{pos}}(t)    \, ,
\label{eq:rdep}
\end{equation}
where $R_{\gamma}$ represents the energy from gamma rays emitted during radioactive decay and can be computed using:
\begin{equation} 
R_{\gamma} = \mathlarger{\mathlarger{\epsilon}_{\mathrm{Ni}}} M_{^{56}\mathrm{Ni}},
\end{equation}
where $\mathlarger{\mathlarger{\mathlarger{\epsilon}_{\mathrm{Ni}}}}$ is a proportionality constant defined in Eq.~\ref{eq:gammaenergy}. The positron energy deposition rate, $R_{\mathrm{dep},\, \mathrm{pos}}(t)$ also depends on the $M_{^{56}\mathrm{Ni}}$ and follows the decay chain of $^{56}$Ni to $^{56}$Co and then to $^{56}$Fe. 
The positron energy deposition rate can be approximated by:
\begin{equation}
R_{\text {pos}}(t) \propto M_{^{56}\mathrm{Ni}} \left(-e^{-t/8.76}+e^{-t/111.4}\right)
\label{eq:rpos}
\end{equation}
At late times (hundreds of days post-peak), positrons can escape the ejecta, making the second term negligible.
As the ejecta expands and becomes nearly transparent, the energy deposited in the ejecta is promptly radiated away. Thus, the bolometric luminosity, $L_{\mathrm{bol}}(t)$, at sufficiently late times (when $t \gg t_{\mathrm{peak}}$) is equal to the energy deposition rate:
\begin{equation}
L_{\mathrm{bol}}(t) = R_{\mathrm{dep}}(t).
\label{eq:bolometric}
\end{equation}
The time-weighted energy in the radiation is conserved due to the predominantly adiabatic expansion of the ejecta. This conservation can be expressed as:
\begin{equation}
    t E_{\mathrm{rad}}(t) + \int_0^t dt' \, t' L_{\mathrm{bol}}(t') = \int_0^t dt' \, t' R_{\mathrm{dep}}(t'),
\end{equation}
where \(E_{\text{rad}}(t)\) is the total energy in UVOIR (ultraviolet, optical, and infrared) radiation within the ejecta at time \(t\). \\
At sufficiently late times, the opacity is low, and the ejecta is almost transparent, resulting in minimal trapped radiation. Thus, for \(t \gg t_{\text{peak}}\):
\begin{equation}
    \int_0^t dt' \, t' L_{\mathrm{bol}}(t') = \int_0^t dt' \, t' R_{\text{dep}}(t').
\end{equation}
Relations for the bolometric luminosity and energy deposition can be used to compare theoretical models with observational data. Particularly, comparing the ratios of these integrals is useful:
\begin{equation}
    \frac{L_{\mathrm{bol}}(t)}{\int_0^t dt' \, t' L_{\mathrm{bol}}(t')} = \frac{R_{\text{dep}}(t)}{\int_0^t dt' \, t' R_{\text{dep}}(t')}.
\end{equation}

This ratio is independent of both distance and the overall $M_{^{56}\mathrm{Ni}}$. By analyzing this ratio from direct observations, the gamma-ray escape time $t_0$ can be estimated. Once $t_0$ is determined, $M_{^{56}\mathrm{Ni}}$ can be inferred by comparing the light curve amplitude to the calculated deposition function using the equations for $R_{\mathrm{dep}}$, $R_{\gamma}$, and $R_{\mathrm{pos}}$.

\section{Results and Discussion}
\label{sec:result}
The results for estimates of $M_{^{56}\mathrm{Ni}}$, based on the methods outlined in Section ~\ref{sec:methodology}, are presented here, accompanied by a comparison of the methods.

\subsection{Estimates Using Arnett's Rule}
\label{sec:result-arnet}
\subsubsection{Correlation between $L_{\mathrm{max}}$ and $t_2$ }
In accordance with Section~\ref{sec:arnett}, a strong correlation exists between the peak bolometric luminosity (\( L_{\text{max}} \)) and the secondary maximum (\( t_2 \)) in the NIR light curve. To test the sufficiency of a linear relationship in describing the dependence of \( L_{\text{max}} \) on \( t_2 \), we fit both the linear (Eq.~\ref{eq:lin}) and the quadratic (Eq.~\ref{eq:cubic}) relations  to 18 SNe Ia data taken from \cite{dhawan2016reddening}. The results of this analysis are presented in Table~\ref{table:abc}. Our analysis shows that the coefficients \( a \) and \( b \) are consistent between the linear and quadratic fits, with the quadratic term \( c \) being negligible. This indicates that the values of \( a \) and \( b \) are not significantly different from those obtained using the linear relation, suggesting that a linear model is sufficient to describe the dependence of \( L_{\text{max}} \) on \( t_2 \).

\begin{table}[htb]
\caption{Comparison of linear and  quadratic fits for estimating \(L_{\text{max}}\) and \(t_2\). The coefficients for both the Y and J filters are: \(a = 0.04\) and \(b = -0.02\).}
\begin{tabular}{@{}lccc@{}}
\toprule
Relation & $a$ & $b$ & $c$\\\midrule
Linear & 0.04 & -0.02 & -\\  
Quadratic &  0.04 & -0.02 & $8.9\times 10^{-14}$  \\ 
\hline
\end{tabular}
\label{table:abc}
\end{table}

$L_{\mathrm{max}}$ is calculated against the phase of second maximum $t_2$. The $t_2$ parameter is a useful tool for calculating the bolometric luminosity of Type Ia supernovae because, in comparison to other parameters utilized in these analyses, it is merely sensitive to reddening and distance. We can take advantage of the coefficients of $a$ and $b$ to get the values of $L_{\mathrm{max}}$ (Eq.~\ref{eq:lin}).
\subsubsection{Deriving $M_{^{56}\mathrm{Ni}}$ from $L_{\mathrm{max}}$}
Building upon Arnett's rule (Section~\ref{sec:arnett}), we estimate the $M_{^{56}\mathrm{Ni}}$ for nine SNe Ia based on their maximum luminosity ($L_{\mathrm{max}}$). Table~\ref{table:mass} presents these mass estimates in two columns. Column 2 utilizes a fixed rise time of $19 \pm 3$ days, while Column 3 leverages the individual rise time calculated for each SNe Ia using the $\Delta m_{15}$ parameter. Interestingly, the fixed rise time approach consistently yields higher $M_{^{56}\mathrm{Ni}}$ values compared to the individual rise time method.
This observed trend highlights the sensitivity of $M_{^{56}\mathrm{Ni}}$ estimates to the adopted rise time. To further illustrate this dependence, Table~\ref{table:fixed} presents mass estimates obtained using fixed rise times of 18, 19, and 20 days. As expected, a longer fixed rise time leads to a systematically larger $M_{^{56}\mathrm{Ni}}$ estimate. Notably, even these estimates remain higher than those obtained using individual rise times.


\subsection{Deriving $M_{^{56}\mathrm{Ni}}$ from bolometric Light Curves}
\label{sec:result-energy}
The luminosity ratio $L_{\mathrm{bol(t)}} / \int_{0}^{t} L_{\mathrm{bol}}(t^{\prime}) t^{\prime} dt^{\prime}$ has been used to extract $t_0$ using Eqs. (13-20). Figure 1 displays the luminosity ratios of nine well-observed supernovae using UV, optical, and Infrared measurements. The luminosity ratios are multiplied by $t^{2.5}$, for convenience, where t is the expected number of days since the explosion. The image describes the light curves of each of the nine supernovae.
Final estimates of Ni mass using the above method are shown in the last column of Table~\ref{table:mass}. It is interesting to note that the values obtained by this method are consistently smaller than those obtained from Arnett's rule.

\begin{figure}[!t]
\includegraphics[width=0.95\columnwidth]{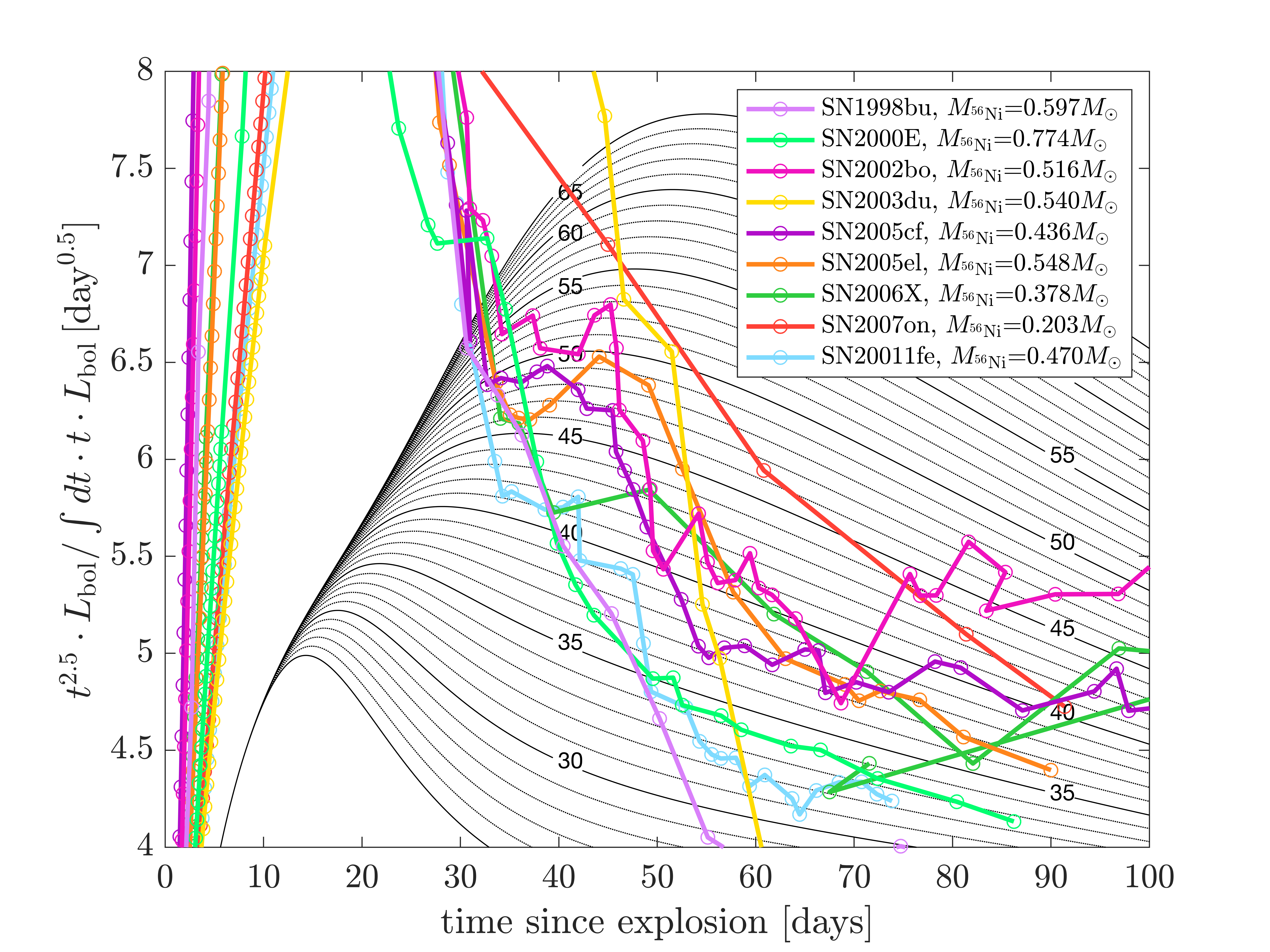}
\caption{The ratio $L_{\mathrm{bol}}(t) / \int {t}{dt} L_{\mathrm{bol}}(t)$ (multiplied by $t^{2.5}$ for convenience) are shown in solid lines 
for a sample of well-observed supernovae.}
\label{fig:energy}
\end{figure}
\begin{table}
\caption{Mass of $^{56}\mathrm{Ni}$ (in units of $M_{\odot}$) using Arnett's rule when different values of fixed Rise Time, $T_r$ (in days) are considered.}
\label{table:fixed}
\begin{tabular}{@{}lccc@{}}
\toprule
SNe & Tr = 18 & Tr = 19 & Tr = 20\\ 
\midrule
SN2005el & 0.474 & 0.498 & 0.522\\ 
SN1998bu & 0.567 & 0.596 & 0.625\\ 
SN2007on & 0.352 & 0.370 & 0.388\\
SN2006X  & 0.536 & 0.563 & 0.590\\
SN2005cf &  0.558 & 0.587 & 0.615\\
SN2003du &  0.590 & 0.620 & 0.650\\
SN2011fe & 0.571 & 0.599 & 0.629\\ 
SN2002bo & 0.549 & 0.577 & 0.605\\
SN2000E & 0.606 & 0.637 & 0.668 \\\hline
\end{tabular}
\end{table}

\begin{table}
\caption{Mass of $^{56}\mathrm{Ni}$ in units of $M_{\odot}$ obtained using Arnett's rule with fixed rise time (19 days) and individual rise time and with energy conservation method.}
\label{table:mass}
\begin{tabular}{@{}lccc@{}}
\toprule
SN  & Fixed rise time & Individual rise time & Energy conservation \\\midrule
SN2005el & 0.498 & 0.440 & 0.548\\ 
SN1998bu & 0.596 & 0.533 & 0.597\\ 
SN2007on & 0.370 & 0.272 & 0.203\\
SN2006X  & 0.563 & 0.533 & 0.378\\
SN2005cf &  0.587 & 0.541 & 0.436\\
SN2003du &  0.620 & 0.587 & 0.540\\
SN2011fe & 0.599 & 0.542 & 0.470\\ 
SN2002bo & 0.577 & 0.566 & 0.516\\
SN2000E & 0.637 & 0.608 & 0.774 \\\hline
\end{tabular}
\end{table}

\subsection{Comparison of the Methods}
\label{sec:result-comparison}
We have employed two different methods to estimate $M_{^{56}\mathrm{Ni}}$. The first method is based on Arnett's rule and there are two variations of this method. The second method is based on energy conservation during the expansion of ejecta. Masses shown in different columns of Table~\ref{table:mass} are different, in fact, Arnett's rule with fixed rise time provides higher values in most of the cases. In this section, we wish to determine if the energy conservation method provides significantly different values of Ni mass than Arnett's rule. We employ the two-tailed student t-test on the paired samples available in Table~\ref{table:mass}. Our null and alternative hypotheses for the energy conservation method vs Arnett's rule are:\\
\begin{itemize}
    \item \textbf{$H_0$}: the mean estimated $M_{^{56}\mathrm{Ni}}$ is the same in two different methods.
    \item \textbf{$H_1$}: The mean estimated $M_{^{56}\mathrm{Ni}}$ from the two different methods is different.
\end{itemize}

To understand if the difference in mean is significance we calculate t-score as $ t = \frac{m_1 - m_2}{\sqrt{s_1/n_1 + s_2/n_2}}$, where $m_1$, $m_2$ are means; $s_1$, $s_2$ are the variances; and $n_1$, $n_2$ are the sizes of the two samples, respectively. It is compared against the critical value which is obtained from the t-test table. A larger t-score compared to the critical value indicates that the difference in means of two samples is significant and hence the null hypothesis is rejected. 
The calculation of the t-statistic between Arnett's rule with individual rise time and energy conservation method is presented in Table~\ref{tab:t-test-1}. The critical value of ``t'' is greater than that of the calculated value, and the $p-$value is smaller than the significance level implying that the t-test failed to reject the null hypothesis. 
Similarly, Table~\ref{tab:t-test-2} presents the t-test result between Fixed rise time and energy conservation method. Again the data failed to reject the null hypothesis. Thus we conclude that there is no significant difference between the mean mass obtained from the two methods.

\begin{table}[htbp]
    \caption{Comparison of $M_{^{56}\mathrm{Ni}}$ obtained using Arnett's rule using individual rise time and energy conservation method. Two-tailed Student's t-test for paired samples has been employed. $m_1$ and $m_2$ are means while $s_1$ and $s_2$ are the variances of the two samples.}
    \begin{tabular}{@{}lc@{}}
    \toprule
    Parameter & Value \\ \midrule
    $m_1$  & 0.514 \\
    $m_2$  & 0.496 \\
    $s_1$  & 0.0104 \\
    $s_2$  & 0.0246 \\
    $t$-score  & 0.285 \\
    $p$-value & 0.629\\
    Degrees of Freedom & 16 \\
    Decision & Failed to reject $H_0$ \\
    \bottomrule
    \end{tabular}
    \label{tab:t-test-1}
\end{table}
\begin{table}[htbp]
    \caption{Comparison of $M_{^{56}\mathrm{Ni}}$ obtained using Arnett's rule using fixed rise time ($19\pm3$ days) and energy conservation method. Two-tailed Student's t-test for paired samples has been employed.}
    \begin{tabular}{@{}lc@{}}
    \toprule
    Parameter & Value  \\ \midrule
    $m_1$  & 0.561 \\
    $m_2$  & 0.496 \\
    $s_1$  & 0.00666 \\
    $s_2$  & 0.0246 \\
    $t$-score  & 1.10 \\
    $p$-value & 0.111 \\
    Degrees of Freedom & 16 \\
    Decision & Failed to reject $H_0$  \\
    \bottomrule
    \end{tabular}
    \label{tab:t-test-2}
\end{table}

\section{Conclusions}
\label{sec:conclu}
This study compares two methods for estimating the mass of $^{56}$Ni produced in Type Ia supernova (SN Ia) explosions. The first method employs Arnett's rule, which posits that peak luminosity is determined by the rate of radioactive decay energy deposition in the ejecta. The second method is based on energy conservation principles. Additionally, we examine the correlation between peak luminosity ($L_{\mathrm{max}}$) and the timing of the secondary peak in near-infrared bands ($t_2$).
Our primary findings are as follows:
\begin{enumerate}
\item The relationship between $L_{\mathrm{max}}$ and $t_2$ is linear. Testing a quadratic relation against the linear model revealed no additional explanatory power, indicating that a linear relationship sufficiently describes the correlation.
\item We compared Arnett's rule (using both fixed and individual rise times) with the energy conservation method. The two methods provide slightly different values of Ni mass, however, Student's t-test analysis shows that the differences are not statistically significant.
\end{enumerate}
These results indicate the robustness of both estimation methods, suggesting their reliability in determining the mass of $^{56}$Ni produced in SN Ia explosions. This consistency across different approaches enhances our confidence in the accuracy of Ni mass estimates, contributing to our understanding of SN Ia physics and its applications in cosmological studies. The slight differences in the nickel mass estimation can be resolved with a larger data set.


\section*{Acknowledgements}
Shashikant Gupta thanks SERB (India) for financial assistance (EMR/2017/003714).

\bibliographystyle{cas-model2-names}

\bibliography{cas-refs}

\end{document}